# Voxel-scale quantum state control in nanorod ensembles using reconfigurable needle beams

G. A. Mantashian[1], D. B. Hayrapetyan*[1,2], P. A. Mantashyan[1,2]

[1]Quantum materials and nanophotonics laboratory, A.B. Nalbandyan Institute of Chemical Physics, 5/2 P. Sevak, Yerevan 0014, Armenia

[2]General Physics and quantum nanostructures department, Russian-Armenian University, 123 H. Emin, Yerevan 0051, Armenia

*Corresponding Author: david.hayrapetyan@rau.am

Precisely addressing single nanostructures inside dense ensembles remains a bottleneck for scalable photonic and quantum-information devices. Here we demonstrate, through comprehensive finite element and variational Monte-Carlo modelling, that a reconfigurable three-dimensional array of needle shaped beams can selectively switch the quantum-optical response of individual *InAs* nanorods embedded in *GaAs*. By tuning the local non resonant intensity pattern the exciton and biexciton energies were calculated, electromagnetically induced transparency (EIT) windows were examined, and correspondingly near-field diffraction carpets were dynamically reshaped. A single parameter—the activation ratio between illuminated and dark nanorods—provides continuous control over photoluminescence peak position (≈ 80 meV) and EIT bandwidth (six times). We further predict fully programmable Talbot self-imaging in nanorod arrays with sub-wavelength pitch. Importantly, the observed Talbot carpets enable spatially resolved identification of which nanorods were excited, offering a powerful diagnostic for verifying structured-light activation schemes. The concept offers a low-crosstalk, wafer-scale route toward reconfigurable quantum emitters, tunable diffractive optics and on-chip slow-light components.

## 1. Introduction

Structures with three-dimensional confinement offer remarkable flexibility and superior optoelectronic properties. Fabrication methods such as molecular beam epitaxy [1,2], chemical vapor deposition [3,4], and lithography [5] allow precise control over size and morphology, enabling tunable properties for a wide array of applications. Beyond conventional uses in LEDs, lasers, and display technologies, such nanostructures play an increasingly critical role in quantum photonics and sensing [6-8].

A standing challenge in this domain is the ability to address individual nanostructures within an ensemble, particularly in systems with certain periodicity. While electrostatic and magnetic fields enable global tuning, they fall short when precise, site-specific control is required. Structured light [9] —spatially and temporally shaped light fields engineered to have specific intensity, phase, and polarization distributions—offers a compelling alternative. It enables controlled interaction with matter at the subwavelength scale, leveraging the spatial degree of freedom of light.

Of particular interest are needle beams [10]—fringe-free, free-space optical Bessel-like beams that represent a specific class of truncated Bessel beams formed under specific conditions [11,12]. Originally introduced by Durnin [13] and later extended through the development of fringes free Bessel-like configurations, these beams combine an ultra-narrow transverse profile with an extended, quasi diffraction-free propagation zone. This allows the central intensity core to remain sharply confined over millimeter- to centimeter-scale distances without the need for continuous refocusing, unlike Gaussian beams. These unique propagation-invariant characteristics have unlocked a broad spectrum of advanced applications, including optical coherence tomography, high-resolution microscopy, laser micromachining, particle manipulation [14-16], etc. Furthermore, recent developments have enabled the generation of near-field and plasmonic needle beams—often referred to as photonic nanojets—with sub-wavelength diameters and extreme aspect ratios ($>10^5$), extending their utility in nanoscale optical probing and manipulation [17-19].

Physically, a needle beam can be viewed as the central lobe of a finite aperture Bessel-like field that arises when a conically phased wavefront interferes along the optical axis. Practical generation routes include axicons [20-22], computer-generated holograms [23] on pixelated spatial light modulators (SLM), and more recently developed phase-encoded dielectric or plasmonic metasurfaces [24,25] that shrink the platform to a sub-millimeter footprint. Because the beam diameter can be held to a few micrometers while the depth of focus exceeds hundreds of Rayleigh lengths, needle beams minimize optical cross-talk and enable layer-by-layer addressing of vertically stacked nanostructures. They also exhibit self-reconstruction [26, 27], allowing the high-intensity core to reform after partial obstruction—a feature that is advantageous when propagating through scattering or inhomogeneous media.

In this work, we present a strategy to address individual nanorods in a three-dimensional periodic ensemble using an array of reconfigurable needle beams. These beams are spatially arranged to form a programmable two-dimensional pattern. Additionally, a standing-wave configuration is employed to modulate the light intensity along the propagation axis [28, 29], enabling selective excitation of vertically stacked nanorod layers. This approach enables volumetric optical addressing with minimal overlap between neighboring beams, providing a platform for high-resolution control of light–matter interactions at the "voxel" scale. The programmability of the beam array allows individual needle beams to be selectively switched, modulated, or repositioned in real time, creating dynamic opportunities for controlling quantum states in complex nanostructured systems.

Moreover, the proposed control system influences the energetic spectra and wave functions of one-particle and few-particle systems confined in nanorods. By manipulating these fundamental properties, we gain control over complex phenomena such as photoluminescence (PL) and electromagnetically induced transparency (EIT). In addition, we predicted fully programmable Talbot self-imaging [30, 31] in nanorod arrays with sub-wavelength pitch. The resulting Talbot carpets provide spatially resolved feedback on which nanorods were optically excited, serving as a powerful diagnostic for validating structured-light activation schemes in dense nanostructured environments. Together, these capabilities lay the groundwork for integrated quantum photonic circuits where each nanostructure acts as a controllable quantum node.

The current paper is structured as follows: in Section 2, we present the physical background and conceptual framework behind the proposed control architecture, including the optical configuration for generating voxel-scale needle beam arrays. Section 3 outlines the theoretical and numerical methods used to model the electronic structure, excitonic interactions, and quantum optical response of individual and ensemble nanorods. In Section 4, we provide a comprehensive analysis of simulation results, focusing on tunable PL, EIT, and near-field diffraction behaviors under various activation scenarios. Special emphasis is placed on the programmable reshaping of optical susceptibility spectra and Talbot self-imaging patterns. Finally, Section 5 concludes the work by summarizing the key outcomes.

## 2. Physical background and conceptual framework

In this work, we consider a three-dimensional (3D) array of cylindrical *InAs* nanorods embedded in a *GaAs* matrix. In our theoretical model, the inter-rod spacing is assumed to be sufficiently large to suppress both electronic tunneling and dipole–dipole interactions, thereby allowing each nanorod to be treated as an isolated quantum emitter (Figure 1). This design reflects recent progress in high-precision nanostructure fabrication, including highly periodic two-dimensional (2D) GaAs nanopillar arrays as demonstrated by Ha et al. [32], where directional lasing was achieved via resonant nanoantennas. Extending this architecture into three dimensions can be realized through planar stacking of lithographically aligned layers, aided by chemical–mechanical polishing and other advanced techniques. Template-assisted growth strategies [33–36] further promise scalable and vertically ordered nanorod arrays.

To selectively control quantum states within this 3D nanorods ensemble, we propose a system based on reconfigurable needle beam arrays capable of volumetric (voxel-scale) optical addressing. As illustrated in Figure 1a, a non-resonant Gaussian laser is expanded and passed through SLM, generating a programmable 2D array of needle beams. These beams propagate through the nanorod array and reflect off a back mirror, forming a standing wave that modulates the optical field along the propagation (Z) axis. The mirror's position can be tuned along the Z-axis, allowing dynamic control over excitation planes within the volume. Additionally, pump and coupling beams are aligned via a polarizing beam splitter (PBS) and uniformly illuminate the array.

This configuration provides full 3D spatial control over the optical excitation profile. Various excitation patterns are illustrated in Figure 1b–f, where transparent cubes denote unexcited nanorods and darker purple cubes represent those illuminated by the needle beams. Two limiting cases are shown: the *natural* state (Figure 1b), where no nanorods are illuminated, and the *homogeneous* state (Figure 1c), where all nanorods are exposed to needle beams.

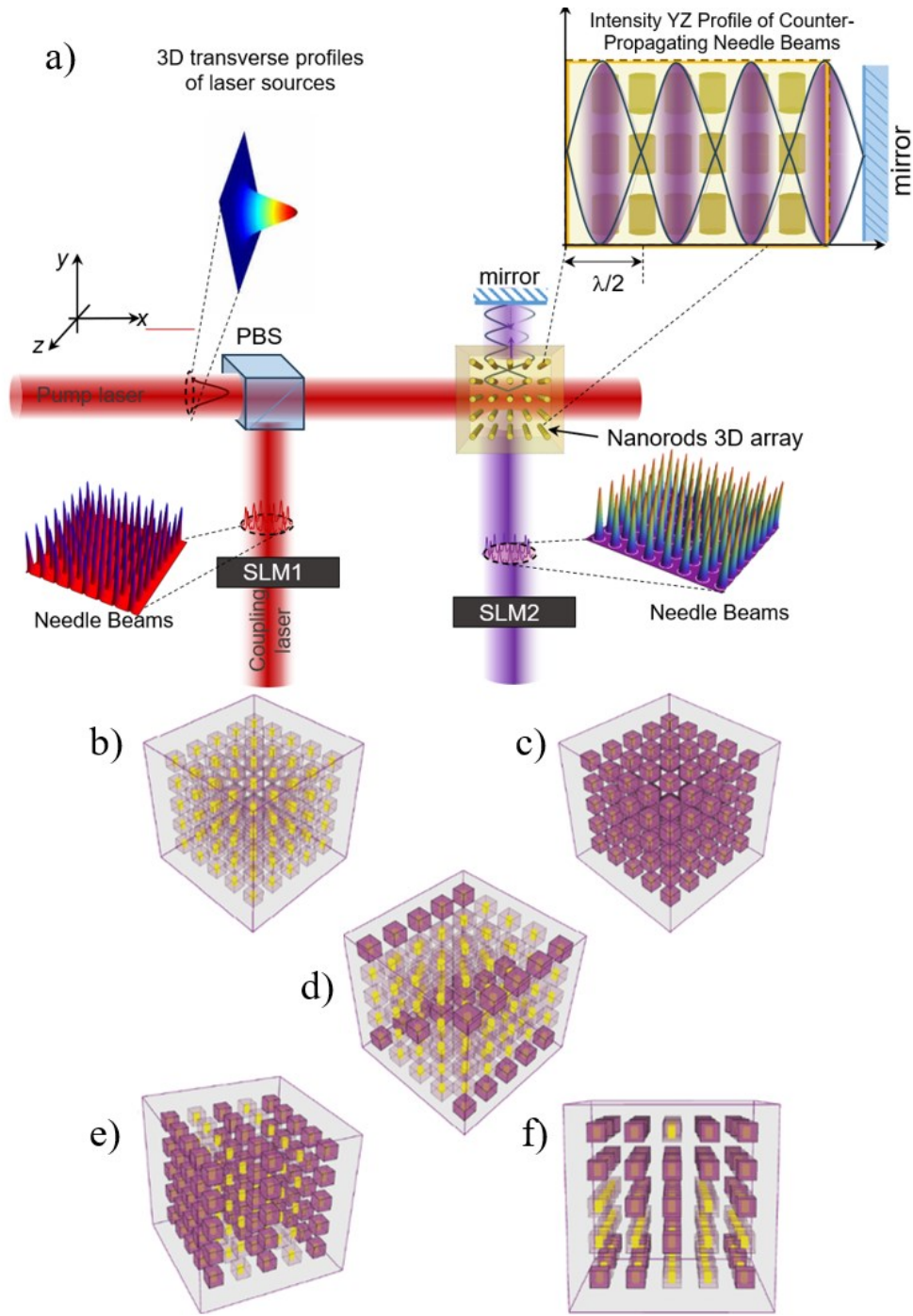

**Figure 1.** (a) Schematic of the proposed voxel-scale nanorod control system. A Gaussian beam is transformed by a spatial light modulator into a 2D array of needle beams that interact with the nanorod array. A back-reflecting mirror forms a standing wave, inducing axial (Z-direction) modulation. The coupling and probe fields irradiate the full volume of nanorods ensemble. *Inset*: intensity distribution in the YZ plane. (b–f) Representative configurations of nanorod excitation states based on the activation ratio. Yellow cylinders represent nanorods. Transparent cubes denote nanorods outside the control field; dark purple indicates actively illuminated nanorods. (b) *Natural* (inactive) state. (c) *Homogeneous* (fully active) state. (e-f) Inhomogeneous configurations with selective activation of lines, planes, or custom patterns.

Between these limits lies a tunable regime of *inhomogeneous* activation (Figure 2c–e), where specific lines, planes, or arbitrary patterns of nanorods are selectively addressed. To quantify this configurational flexibility, we define a single parameter — the activation ratio, denoting the fraction of nanorods actively illuminated by needle beams relative to the total. Given the non-interacting nature of the nanorods and minimal cross-talk due to the needle beams' confinement, this parameter is sufficient to characterize the optical control regime. Beyond localized excitation and energy-level modulation, our control method enables manipulation of the collective optical response of nanorod ensembles. In particular, we examine how a single layer (2D plane) of nanorods behaves as a diffraction grating for the probe laser field, and how this response varies under different spatial configurations and activation regimes.

Assuming that the probe field is a uniform plane wave incident perpendicular to the vertical cross-sections of the nanorods, each nanorod can be approximated as a rectangular aperture. This allows us to treat the entire nanorod layer as a rectangular-aperture-based amplitude grating. The modulation of the probe field is governed by the local optical susceptibility of each nanorod, which is dynamically altered through EIT when the coupling field is strong. Thus, the *activation pattern* indirectly controls the transmission amplitude and phase, forming a programmable diffraction grating.

## 3. Theoretical and numerical methods

To model the quantum optical response of individual and ensemble *InAs* nanorods under needle beams control, we adopted a multiscale computational approach combining analytical transformations, numerical eigenstate calculations, variational quantum methods, and open quantum system dynamics. First, the single-particle electron and hole eigenstates were computed within the framework of the envelope function approximation and single-band effective mass theory, incorporating the Kramers–Henneberger transformation to account for intense laser field (ILF) dressing [37]. A Woods–Saxon potential [38] was used to model the gradual material transition between the *InAs* nanorods and the *GaAs* matrix, including a tunable diffusion parameter. These eigenstates were computed using the finite element method (FEM) implemented in *Wolfram Mathematica* [39]. Full derivations and parameters are provided in Supplementary Section S1.

Using the obtained wavefunctions and energies, we constructed exciton and biexciton states via the Variational Quantum Monte Carlo Method [40]. A screened Coulomb potential was modified by ILF dressing to derive the interaction terms. We employed correlated Gaussian–Jastrow trial wavefunctions for the exciton case [41] and a minimal-parameter biexciton trial function derived from Takagahara's approach [42]. Ground state energies were estimated using adaptive quasi-Monte Carlo integration with parameter optimization by energy minimization (see Supplementary Section S2 for full algorithmic details and implementation notes).

The quantum optical response, including PL and EIT, was modeled by solving a Lindblad master equation [44] for a three-level quantum system representing the ground, exciton, and biexciton states. Optical susceptibility spectra were obtained under steady-state conditions and used to compute absorption and emission via the Roosbroeck–Shockley relation. Supplementary Section S3 includes detailed derivations of the absorption coefficient, emission spectrum, and susceptibility tensor components.

Lastly, to analyze near-field diffraction and Talbot self-imaging, the structured nanorod arrays were modeled as programmable amplitude and phase gratings, defined by the activation patterns of the needle beams. Fresnel diffraction was computed using Fourier-based propagation methods [44] under the paraxial approximation, with structured gratings represented in both real and spatial frequency domains. This allowed us to efficiently simulate the evolution of structured diffraction carpets across various control states (see Supplementary Section S4 for the grating definitions and numerical methods). All numerical simulations were validated through convergence tests and were repeated for multiple beam configurations and activation ratios to ensure reproducibility and robustness.

## 4. Results and discussion
### 4.1. Optical Susceptibility in a Ladder-Type Nanorod System

We begin our analysis by examining the optical susceptibility $\chi$ of the three-level nanorod system under various excitation and control configurations. The optical susceptibility serves as a key indicator of the strength and nature of light–matter interactions, with its imaginary part $\text{Im}(\chi) \approx \alpha(\hbar\omega_p)$ corresponding to the absorption of the probe field and its real part $\text{Re}(\chi) \approx n(\hbar\omega_p)$ governing modulation of the refractive index. These two components collectively determine the optical transparency and dispersion characteristics of the nanorod ensemble, especially under conditions of coherent excitation including EIT. For the specific system under consideration modeled as a ladder-type three-level quantum system analytical expressions for $\chi$ were derived under steady-state conditions by solving the Lindblad master equation (see Supplementary Material, Section S3). These expressions form the basis for simulating the system's absorption and dispersion responses as functions of the pump energy, coupling strength, and nanorod activation pattern, which we now explore in detail.

$$Im\,\chi = \frac{2n_{NR}|\mu_{X0}|^2}{\varepsilon_0 \hbar}$$
$$\times \frac{\gamma_{XX0}\left[\Omega_c^2 + \gamma_{X0}\gamma_{XX0} - \Delta_p(\Delta_p + \Delta_c)\right] + (\Delta_p + \Delta_c)\left[\gamma_{X0}(\Delta_p + \Delta_c) + \gamma_{XX0}\Delta_p\right]}{\left[\Omega_c^2 + \gamma_{X0}\gamma_{XX0} - \Delta_p(\Delta_p + \Delta_c)\right]^2 + \left[\gamma_{X0}(\Delta_p + \Delta_c) + \gamma_{XX0}\Delta_p\right]^2} \quad (1)$$

$$\text{Re}(\chi) = \frac{2n_{NR}|\mu_{X0}|^2}{\varepsilon_0 \hbar}$$
$$\times \frac{-(\Delta_p + \Delta_c)\left[\Omega_c^2 + \gamma_{X0}\gamma_{XX0} - \Delta_p(\Delta_p + \Delta_c)\right] + \gamma_{XX0}\left[\gamma_{X0}(\Delta_p + \Delta_c) + \gamma_{XX0}\Delta_p\right]}{\left[\Omega_c^2 + \gamma_{X0}\gamma_{XX0} - \Delta_p(\Delta_p + \Delta_c)\right]^2 + \left[\gamma_{X0}(\Delta_p + \Delta_c) + \gamma_{XX0}\Delta_p\right]^2} \quad (2)$$

where $n_{NR}$ is the number of illuminated nanorods, $\mu_{X0}$ - the dipole matrix element between the exciton and ground states, $\varepsilon_0$ - *InAs* dielectric constant, $\gamma_{X0}, \gamma_{XX0}$ - are the decay rates for the exciton $|X\rangle$ to ground $|0\rangle$ and biexciton $|XX\rangle$ to exciton $|X\rangle$ transitions, respectively. $\Omega_p, \Omega_c$ are the Rabi frequencies of the corresponding transitions. The detuning parameters $\Delta_p$ and $\Delta_c$ correspond to the frequency offsets of the probe and coupling fields from their respective resonant transitions.

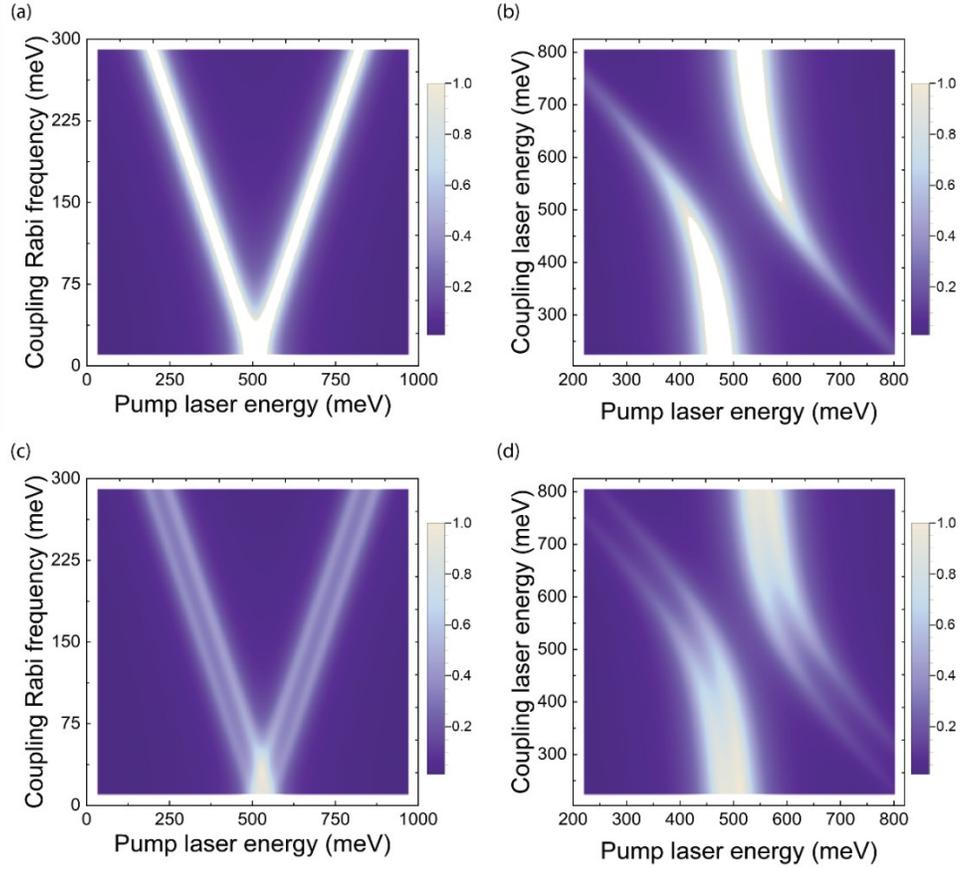

**Figure 2.** Density plots of the imaginary part of the optical susceptibility χ, proportional to the absorption coefficient of the probe field. In all plots, the X-axis denotes the energy of the pump laser $\hbar\omega_p$. (a) and (c) Y-axis represents the coupling Rabi frequency at fixed near-resonant $\hbar\omega_c \approx 495$, $\Delta_c = 0$. (b) and (d) Y-axis shows the coupling Rabi frequency, and the coupling laser energy is fixed at $\Omega_c = 80\ meV$. (a) and (b) correspond to the *natural* state where no nanorods are illuminated. (c) and (d) correspond to the case where 80% of the nanorods are activated using structured needle beams with intensity $4.58364 \cdot 10^{13}\ W/cm^2$.

All physical and material parameters used in the simulations are summarized in Table S1 (Supplementary Information). In Figures 2a and 2c the density plots for the imaginary part of the optical susceptibility χ are presented for the *natural* state, where no nanorods are illuminated by the control (needle beam) system (Figure 1b). Figure 2a maps the imaginary part of the optical susceptibility $\chi$ which is related to the absorption response as a function of the pump laser energy (X-axis) and the coupling field Rabi frequency $\Omega_c$ (Y-axis), with the coupling laser energy fixed at its resonant value $\Delta_c = 0, \hbar\omega_c = 495\, meV$. At low values of $\Omega_c$, the absorption spectrum displays a single broad peak centered at the exciton resonance, indicating negligible quantum interference between states. As the Rabi frequency increases, starting from ~35 $meV$ ( $\Delta_p = 0, \hbar\omega_p = 501\, meV$ ) we observe the emergence of a narrow transparency window flanked by two absorption peaks which is a characteristic signature of EIT. This transparency window arises from destructive interference between the excitation pathways involving the probe and coupling fields. Notably, the width of the EIT window expands with increasing $\Omega_c$, illustrating the enhanced coherence and stronger coupling between quantum states under more intense coupling fields.

Let us move on to analyzing the case of Figure 2b, here the $\Omega_c$ is fixed at 80 $meV$ mark, however, the Y-axis denotes the coupling laser energy $\hbar\omega_c$. Unlike the previous case where the absorption exhibited a V-shape here, it exhibits an N-shaped spectrum. Here we can see clearly that the case where two peaks in the probe absorption are only equal around the coupling resonance. At $\Delta_c = 0$ maximum quantum coherence is achieved leading to effective destructive interference. When the detuning is different from zero, we can see that one of the absorption peaks begins to dominate the other, moreover, as the detuning increases, we see an additional increase in the transparency window.

We now turn to the case in Figures 2c and 2d, where the control system is activated, with 80% of the nanorods illuminated by structured needle beams at intensity $I_{NB} = 4.58364 \cdot 10^{13}\, W/cm^2$. This specific activation level was chosen because it clearly reveals the effects of localized field-induced modulation. For this case, the coupling laser energy is taken as an intermediate value $\hbar\omega_c = 517.5\, meV$, averaging the coupling resonances of the inactive (*natural*) and active (*controlled*) nanorod populations. This results in significant spectral modifications. First, the resonance energies shift, reflecting the influence of intense laser field dressing on the energy levels, $\Delta_p = 0, \hbar\omega_p = 542\, meV$, $\Delta_c = 0, \hbar\omega_c = 540\, meV$. Moreover, we observe new spectral features emerging: for low than ~45 $meV$ Rabi frequencies, a bright triangular central peak appears, accompanied by two weaker satellite peaks on either side. As the coupling Rabi frequency increases, a second set of satellites becomes visible, and the central peak begins to diminish—a signature of the emergence of a transparency window between the two absorption doublets. Each doublet exhibits a local minimum, indicating partial suppression of absorption, though these regions are not fully transparent, as will become more evident in Figure 3. In the final plot of Figure 2d we can see that analogously to the case discussed previously we have an emergence of the second set of peaks in the absorption and sure enough the spot where these peaks are all relative is around $\hbar\omega_c \approx 517.5\, meV$.

*4.2. EIT under different conditions of illumination*

To further explore the EIT phenomenon in the nanorod ensemble, Figure 3 presents the dependence of the imaginary $\text{Im}(\chi)$ and real $\text{Re}(\chi)$ parts of the optical susceptibility $\chi$ on the pump laser energy for varying coupling Rabi frequencies $\Omega_c$. Each row in the figure represents a different activation scenario: the *natural* state with no needle beam illumination (Figure 3a,b); full *homogeneous* activation of the nanorod ensemble by needle beams with intensity values $I_{NB} = 5.09 \cdot 10^{12} \, W/cm^2$ (Figure 3c,d) and $I_{NB} = 4.58 \cdot 10^{13} \, W/cm^2$ (Figure 3e,f).

In the unperturbed, or *natural*, configuration (Figures 3a, b), the nanorods are not illuminated by any control field. At $\Omega_c = 0$, the coupling field is off, the coupling laser is turned off, and the system exhibits a single absorption peak centered at $\Delta_p = 0$, corresponding to the pump resonance energy of 502 meV. This peak results from direct transitions between the ground and exciton states, without quantum interference. The real part of the susceptibility shows a sharp normal-to-anomalous dispersion transition near resonance, with the refractive index decreasing as the pump energy increases. As the coupling Rabi frequency increases, a clear transparency window emerges between two absorption peaks—a typical signature of EIT. This spectral splitting arises from destructive interference between two excitation pathways in the ladder-type three-level system (see Supplementary Section S3). In the corresponding $\text{Re}(\chi)$ profile (Figure 3b), we observe the development of a structured dispersion curve with triple zero-crossings, indicative of sharp refractive index modulation near the transparency window.

When the nanorod array is *homogeneously* illuminated with moderate intensity $I_{NB} = 5.09 \cdot 10^{12} \, W/cm^2$ (Figures 3c, d), the qualitative features of EIT are retained, but the spectral positions shift due to changes in the local electronic environment. Specifically, the pump and coupling resonances are now centered at 521 meV and 517 meV, respectively. These shifts reflect intense laser field induced renormalization of the energy levels. The absorption peaks also become asymmetric, with the left-side peaks (lower energy) gaining relative intensity over the right-side peaks. This asymmetry is most pronounced at lower coupling strengths (e.g., $\Omega_c = 35$ meV), where coherence is more sensitive to detuning. The real part of the susceptibility likewise becomes asymmetric, and for $\Omega_c = 35$ meV, the zero-crossings are suppressed entirely, with the dispersion remaining negative across the resonance. This indicates that the transparency is less pronounced, and coherence is reduced. The overall magnitude of both $\text{Im}(\chi)$ and real $\text{Re}(\chi)$ decreases by ~20% compared to the natural state, which can be attributed to differential responses of electrons and holes to the intense laser field, given their mass disparity (see Table S1 and Figure S1 in Supplementary Information). This discrepancy leads to a reduction in the effective dipole moment and consequently weaker optical transitions.

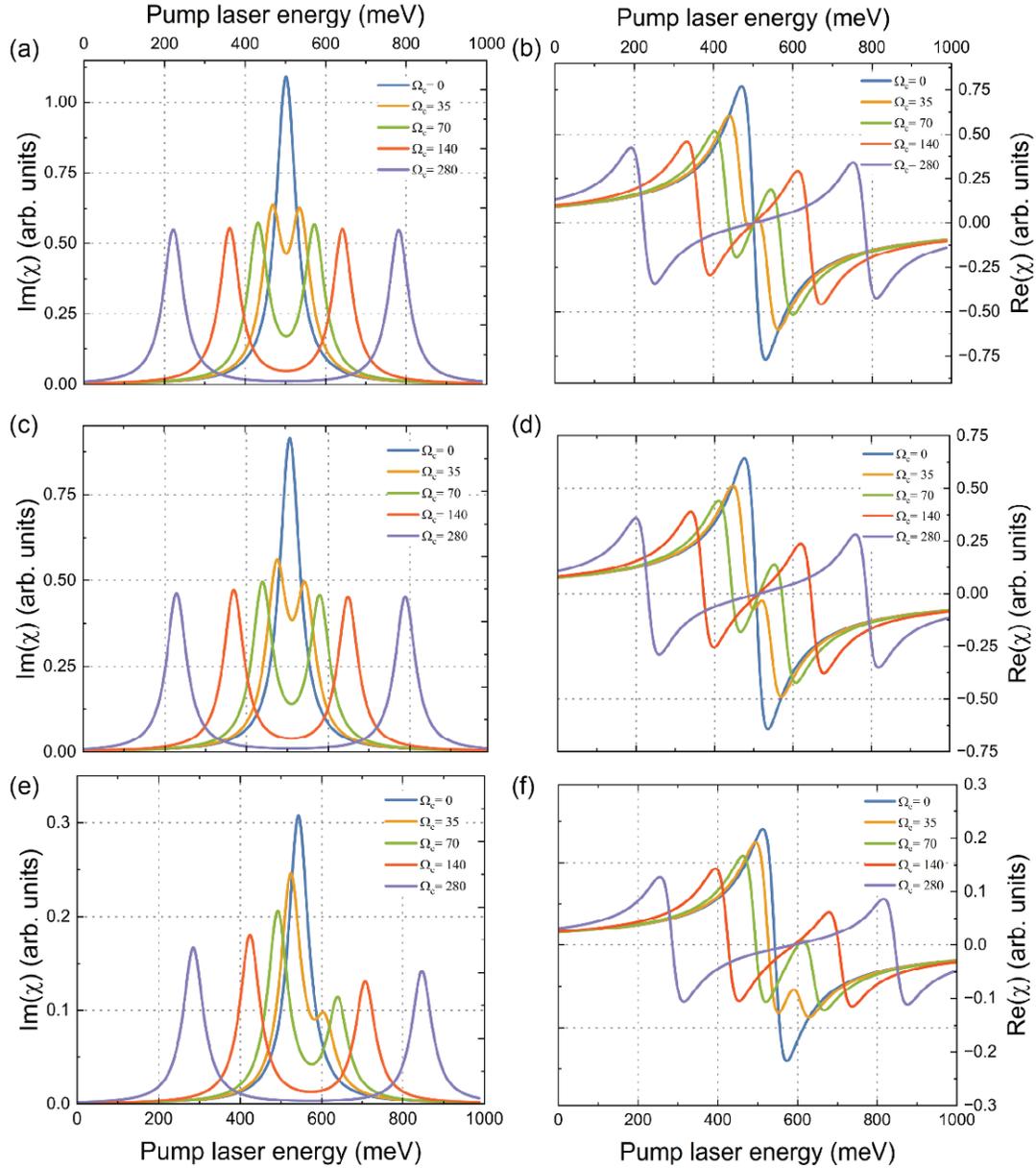

**Figure 3.** Dependence of the imaginary (left column) and real (right column) parts of the optical susceptibility χ on the pump laser energy for different values of the coupling Rabi frequency $\Omega_c$, under varying nanorod activation conditions. (a, b) Natural state with no nanorods illuminated $I_{NB} = 0$. (c, d) Homogeneous activation of the nanorod ensemble (100% of nanorods are illuminated $I_{NB} = 5.09 \cdot 10^{12}\, W/cm^2$). (e, f) Homogeneous activation using structured needle beams at an intensity of $I_{NB} = 4.58 \cdot 10^{13}\, W/cm^2$. Each row corresponds to a different control configuration, while the coupling strength is varied within each subplot to highlight the evolution of absorption and refractive index modulation across regimes.

In the final set of plots, we can see where we see that for the case where the intensity of the control system is increased further to the $I_{NB} = 4.58 \cdot 10^{13}\, W/cm^2$ (Figures 5d,f) we can see that the changes in both imaginary and real parts of the optical susceptibility are enhanced further, and now the difference in peak amplitudes is visible even in the case of $\Omega_c = 280 meV$. Moreover, the zero crossing points in the dispersion of $\Omega_c = 70 meV$ have been pushed down further and for the $\Omega_c = 35 meV$ case, the oscillatory region is deeper in the negative region. All of these features demonstrate the way in which the properties of the EIT spectrum can be controlled by employing the control system proposed in the current article. Notably, we can say that in the case when we change the $\hbar\omega_c$ to correspond to the $\Delta_c$ for each laser intensity, we would see that the relative peak magnitudes stay equal to each other and we get a simple change in the peak and transparency window positions.

### 4.3. Effect of full and structured activation on the system optical response

Finally, we turn to the results illustrating the impact of inhomogeneous activation of nanorods on both the PL response (Figure 4a) and the imaginary part of the optical susceptibility $Im(\chi)$, which is directly related to probe absorption (Figure 4b), for various activation ratios at a fixed control field intensity $I_{NB} = 4.58 \cdot 10^{13} W/cm^2$. The activation ratio $n_{A:iA}$ refers to the percentage of active $n_A$ and inactive $n_{iA}$ nanorods within the ensemble, controlled via the structured needle beam illumination.

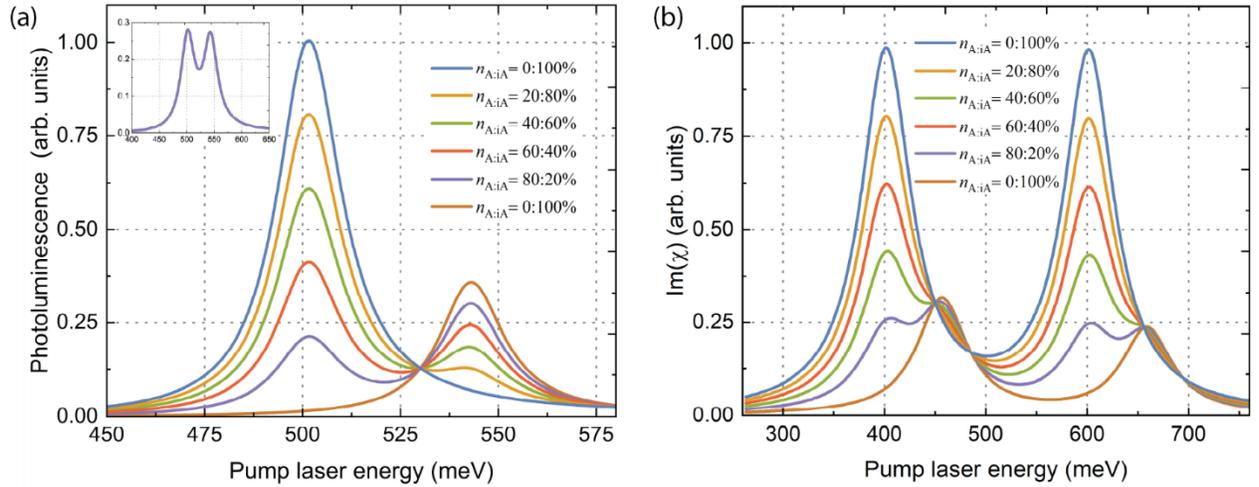

**Figure 4.** (a) PL between the $|0\rangle$ and $|X\rangle$ levels of the nanorod ensemble as a function of pump laser energy for various activation ratios $n_{A:iA}$, at fixed control field intensity of $4.58 \cdot 10^{13} W/cm^2$ and the inhomogeneous line broadening $\Gamma = 15 meV$. The *inset* highlights the crossover point around $n_{A:iA} = 74:26\%$ where both peaks exhibit equal intensity. (b) Imaginary part of the optical susceptibility $Im(\chi)$ on the pump laser energy under the same activation ratios and control intensity.

It can be seen from Figure 3a that as the proportion of activated nanorods increases, the PL spectrum exhibits a clear evolution from a single-peaked Lorentzian profile at $n_{A:iA} = 0:100\%$ (*natural* state) to a pronounced double-peaked structure. This spectral splitting arises due to the coexistence of nanorods with different energy resonances—those illuminated by the control field and those left in their unperturbed state. The inset in Figure 4a highlights the crossover point, where both peaks attain equal intensity at around $n_{A:iA} = 74:26\%$ demonstrating that the ensemble emission becomes strongly bimodal under balanced activation conditions. At full activation $n_{A:iA} = 100:0\%$, the PL peak entirely shifts to the higher-energy side, corresponding to the modified resonance of the controlled nanorods. The picture is slightly different in the case of $\text{Im}(\chi)$ (Figure 4(b)), we can see that the gradual shift visible for the PL does not happen for EIT. We see that the quartet of peaks visible in Figure 2(c) is discernable only for the $n_{A:iA} = 60:40\%$ and $n_{A:iA} = 80:20\%$. However, as the EIT is a much more complex phenomena that is dependent on many other parameters it is hard to pinpoint the exact ratio for the regime where all 4 peaks are equal.

Together, these results confirm that partial activation of nanorod ensembles allows for tunable optical responses—both in spontaneous emission and probe absorption—through spectral reshaping driven by coherent control. This tunability is essential for implementing spatially programmable optical functionalities, such as reconfigurable diffraction gratings or multiplexed quantum light sources.

### 4.4. Observation of near-field diffraction

In this section, we investigate the near-field diffraction behavior of a probe laser field transmitted through 2D nanorod ensemble, under various spatial configurations and activation patterns (Figure 1). When illuminated by a resonant coupling laser field capable of inducing EIT, the nanorods array behaves as an optically reconfigurable diffraction grating. This optical gating mechanism enables modulation of the amplitude and phase of the incident probe wave, effectively controlling the resulting diffraction landscape. Assuming the probe beam is a normally incident plane wave interacting with the vertical cross-section of the nanorod array, the illuminated nanorods are modeled as rectangular apertures, and the overall diffraction pattern is governed by their distribution, orientation, fill factor, and activation state. All simulations account for full 2D grating effects, and the central X–Z and Y–Z slices of the resulting intensity distributions are presented to illustrate the near-field diffraction behavior.

#### 4.4.1. Near-field diffraction in disordered nanorod arrays

We begin by analyzing a disordered ensemble of nanorods distributed randomly in a 2D plane, with both positional and orientational randomness (Figure 5a). Under uniform plane-wave coupling at a sufficiently high Rabi frequency $\Omega_c$ to induce transparency, the activated nanorods act as amplitude modulation elements for the probe field. The resulting intensity distributions in both the X–Z and Y–Z planes (Figures 5b and 5c) display speckle-like, diffuse scattering, with no discernible periodic features. This behavior is characteristic of random phase/amplitude gratings and highlights the absence of spatial coherence in the nanorod layout.

Next, we consider a partially ordered case (Figure 5d), where nanorods are uniformly aligned along the Y-axis and arranged in a single monolayer without overlapping. Such ordering is experimentally achievable via directed growth techniques including hydrothermal synthesis or seed-assisted deposition. Despite improved alignment, the resulting diffraction pattern (Figures 5e and 5f) remains largely diffuse and irregular, indicating that orientational uniformity alone is insufficient to induce well-defined periodic diffraction under random spatial placement.

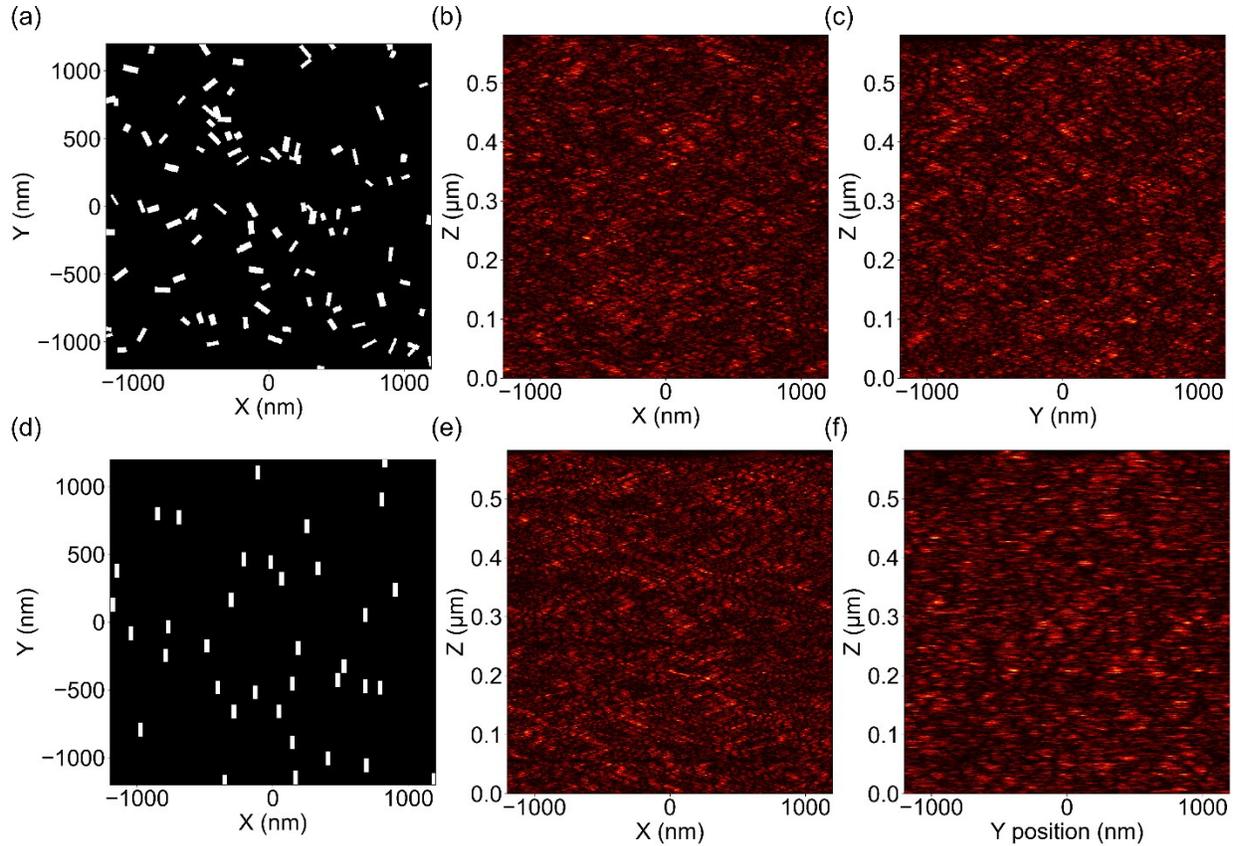

**Figure 5.** Near-field diffraction of the probe field through disordered nanorod distributions. (a) Schematic of a 2D nanorod ensemble with completely random spatial distribution and orientation. (b, c) Simulated probe intensity profiles in the X–Z and Y–Z planes, respectively, for a uniformly illuminated grating with high coupling Rabi frequency $\Omega_c$. (d) Schematic of a partially ordered nanorod array with all rods aligned along the Y-axis and positioned in a single non-overlapping monolayer. (e, f) Probe field intensity distributions in the X–Z and Y–Z planes.

*4.4.2. Periodically Positioned Nanorods with Random Orientation*

To enhance spatial coherence, we examine nanorod arrays with periodic positioning but random orientation, as shown in Figure 6. In this case, we can switch our suggested setup (Figure 1) to the one where our coupling laser beam is transformed into a 2D array of needle beams with an arbitrary pattern. Using this configuration, we can address individual nanorods and obtain various random distributions. Here, the nanorod centers are located on a square grid with a fixed pitch of either 250 nm or 400 nm, while the rods are randomly oriented along the X- or Y-axis. The probe diffraction patterns (Figures 6b–c and 6e–f) reveal a transition toward partial periodicity, as evidenced by the emergence of structured interference fringes amidst background speckle noise. Notably, increasing the grating period to 400 nm reduces the intensity of diffuse features and promotes Talbot-like carpet formation, although residual irregularities persist due to orientation disorder.

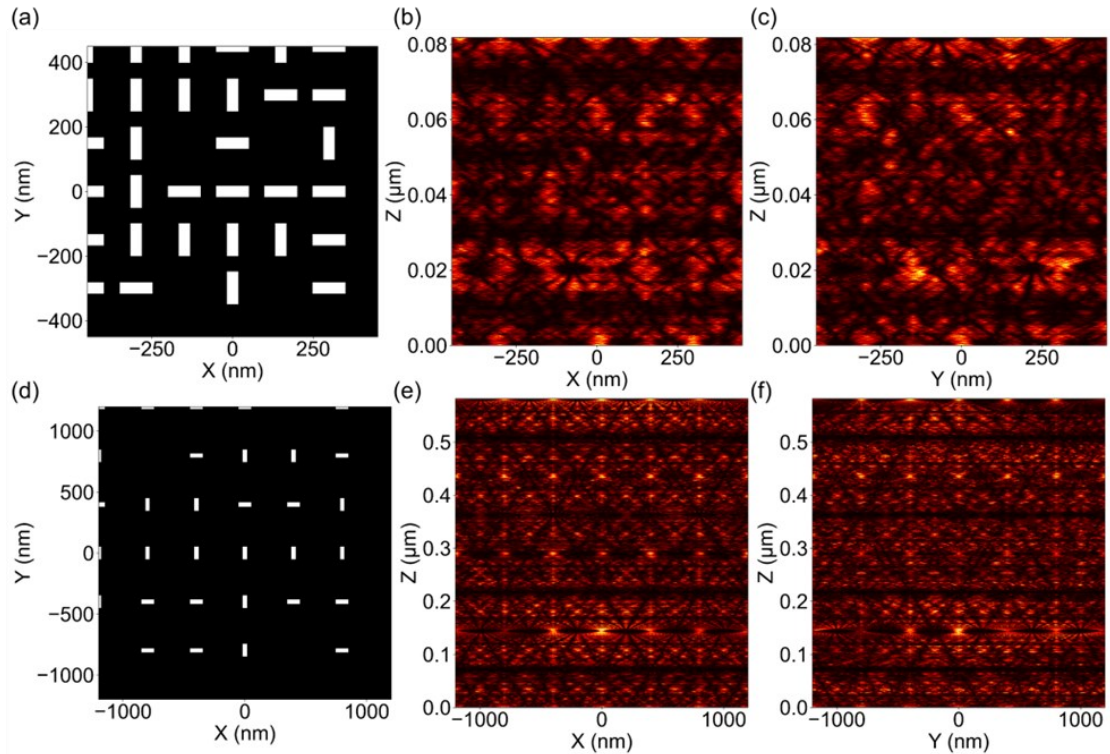

**Figure 6.** Probe field diffraction through structured 2D nanorod arrays with positional periodicity and random orientation. (a) Schematic of nanorods positioned on a 2D periodic square grid with 250 nm pitch, but randomly oriented along the X- or Y-axis. (b, c) Corresponding X–Z and Y–Z probe field intensity distributions. (d) Similar array with increased grating period of 400 nm. (e, f) Corresponding X–Z and Y–Z probe field intensity distributions.

### 4.4.3. Influence of Fill Factor on Periodic Gratings

In Figure 7, we show the impact of fill factor, defined as the fraction of grid positions occupied by nanorods, on near-field diffraction in rectangular-aperture-based periodic gratings. Here, nanorods are consistently aligned along the Y-axis and positioned on a 2D lattice with 150 nm spacing. As the fill factor increases from sparse (Figure 9a–c) to dense (Figure 9j–l), the diffraction pattern evolves from highly noisy and fragmented to increasingly structured and periodic. Even at low fill factors, faint periodic features begin to emerge; however, destructive interference between randomly spaced apertures generates significant distortion. At higher fill factors, this interference is gradually suppressed, and Talbot-like carpets begin to dominate the intensity distributions. Nevertheless, complete suppression of diffuse noise is not achieved even at maximum fill, indicating persistent sensitivity to residual structural disorder.

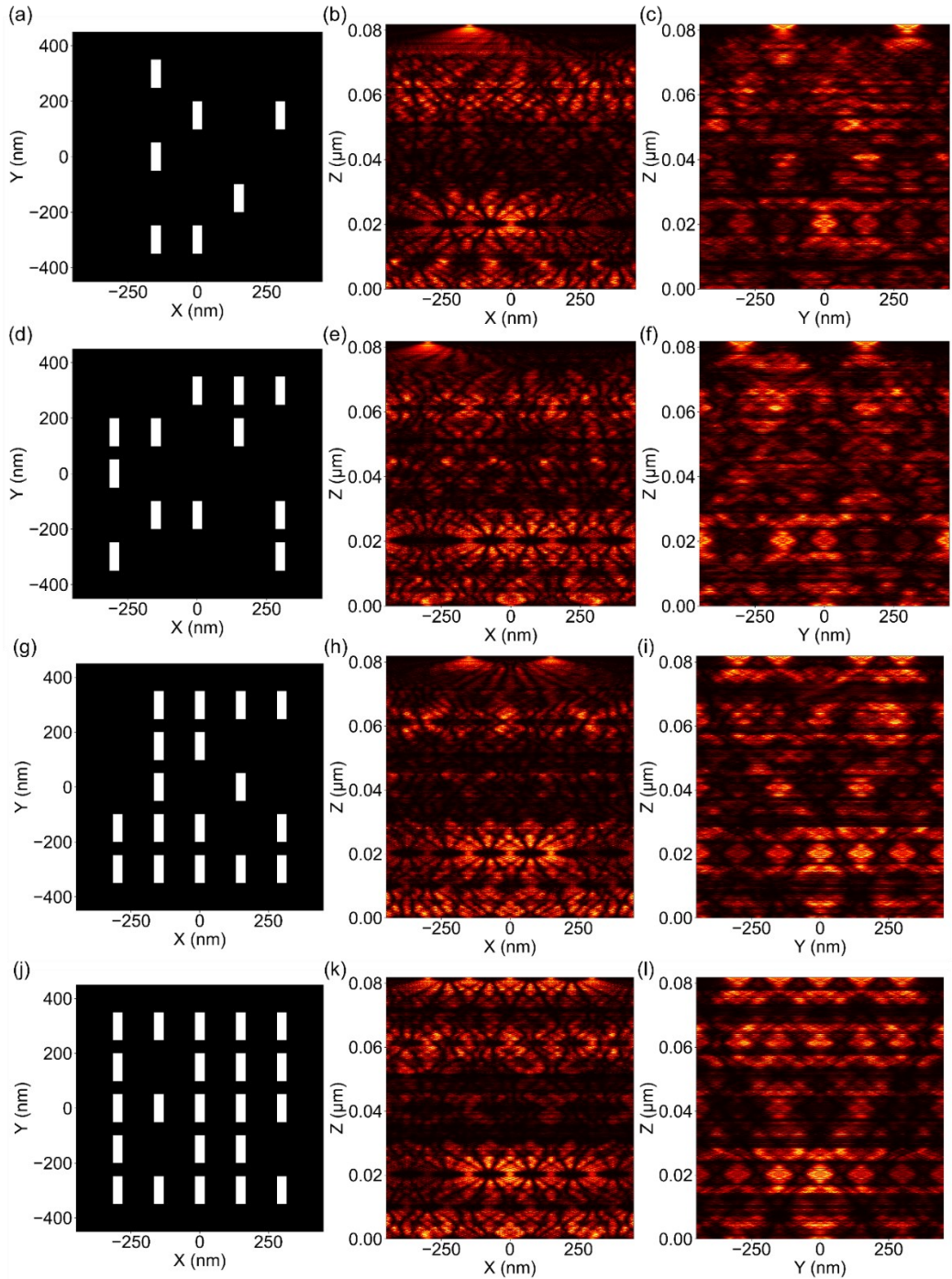

**Figure 7.** Effect of increasing fill factor in periodic nanorod-based random gratings on near-field probe diffraction patterns. (a, d, g, j) Schematics of rectangular apertures formed by nanorods aligned along the Y-axis and placed on a periodic grid with 150 nm spacing; the fill factor increases from top to bottom. (b, e, h, k) Corresponding X–Z and (c, f, i, l) Y–Z probe field intensity distributions.

*4.4.4. Fully Periodic Nanorod Arrays and Talbot Self-Imaging*

To benchmark ideal grating performance, Figure 8 presents diffraction from fully periodic nanorod arrays with complete activation of all apertures on grids of 150 nm and 400 nm spacing. In these idealized cases, the resulting X–Z and Y–Z intensity distributions (Figures 8b–c and 8e–f) exhibit highly regular Talbot patterns, consistent with self-imaging behavior expected from periodic diffractive structures. The anisotropic shape of the nanorods (rectangular apertures) leads to distinct Talbot carpets in the X–Z and Y–Z planes. Importantly, this configuration confirms that structured control of activation combined with spatial periodicity enables reconfigurable near-field diffraction with high spatial coherence, tunable via the grating period and nanorod orientation.

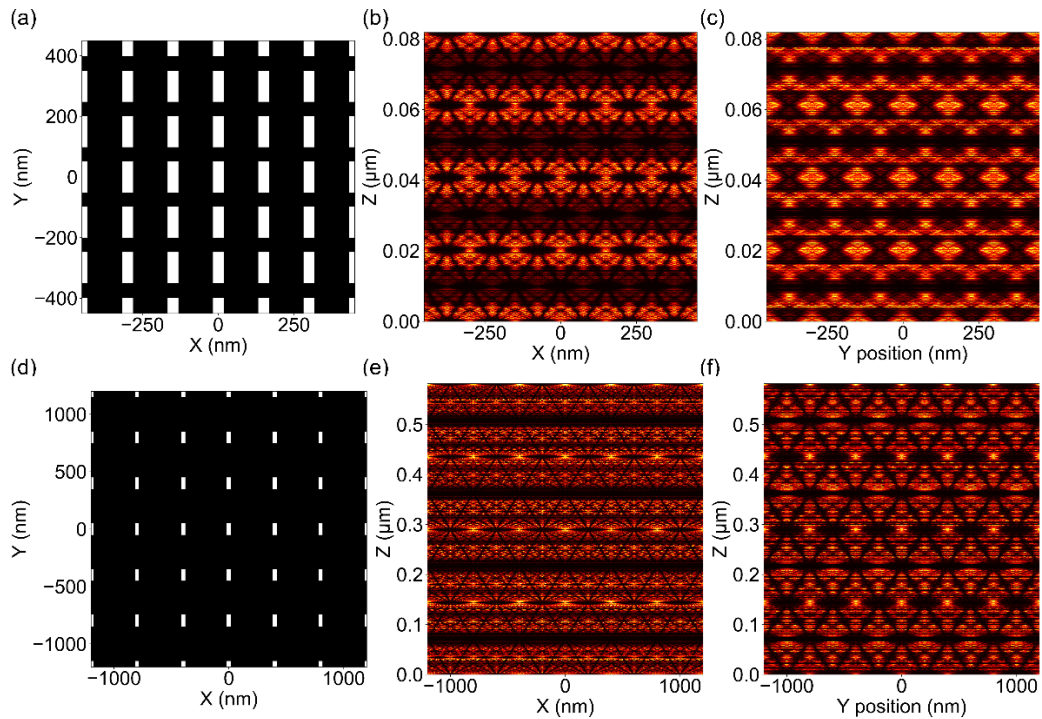

**Figure 8.** *Fully periodic nanorod arrays with complete aperture activation demonstrating Talbot self-imaging.* (a, d) Schematics of gratings where all grid points contain nanorods aligned along the Y-axis, with periods of 150 nm and 400 nm, respectively. (b, e) Corresponding X–Z probe field intensity distributions, and (c, f) Y–Z probe field intensity distributions.

*4.4.5. Line-Selective Activation and Anisotropic Grating Control*

Expanding on this concept, Figure 9 demonstrates the effect of selective row and column activation, forming triple-line gratings with vertical or horizontal orientation at 150 nm (Figure 9a) and 400 nm periods (Figure 9d). Due to the rectangular geometry of the nanorods, horizontal and vertical line gratings produce distinct interference patterns (figure 9b,c and e.f). The Y–Z slices in the horizontal configuration (fewer apertures in the central plane) exhibit more diffuse intensity compared to the X–Z slices of the vertical configuration, which include more immediately contributing apertures. These observations confirm that the local aperture density within a given plane strongly influences the sharpness and complexity of the resulting near-field pattern, even when full 2D integration is performed in simulations.

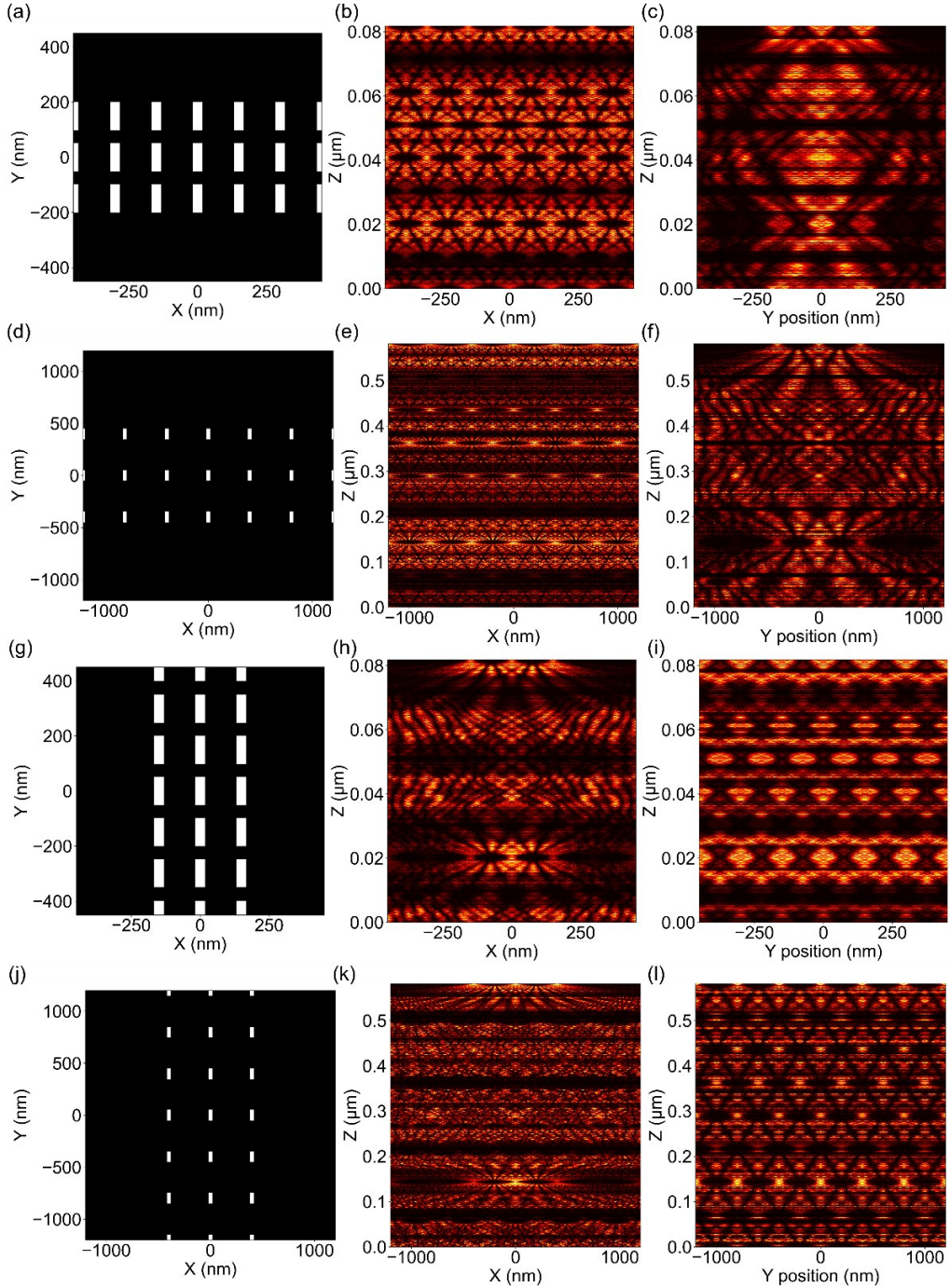

**Figure 11.** Near field diffraction from selective nanorods line activation illustrating the influence of aperture orientation, count, and spacing. (a, d) Grating layouts with three horizontal nanorod lines at 150 nm and 400 nm spacing, respectively. (b, e) Corresponding X–Z intensity distributions. (c, f) Corresponding Y–Z intensity distributions. (g, j) Grating layouts with three vertical nanorods lines at 150 nm and 400 nm spacing. (h, k) Corresponding X–Z intensity distributions. (i, l) Corresponding Y–Z intensity distributions.

*4.4.6. Cross-Diagonal Gratings and Diagonal Talbot Patterns*

Finally, in Figure 10, we explore cross-diagonal grating arrangements, where nanorods are activated along two diagonals intersecting at ±45° with respect to the X and Y axes. At a small grating period of 150 nm (Figures 10a–c), the resulting diffraction patterns display rich diagonal interference fringes interwoven with speckle noise due to high spatial frequency and aperture proximity. Upon increasing the grating pitch to 400 nm (Figures 10d–f), the diffraction pattern becomes more coherent and structured, with emergent diagonal Talbot carpets and improved periodicity. The appearance of multiple self-replicating interference fringes aligned with the diagonal activation axes reveals the unique symmetry introduced by this grating configuration.

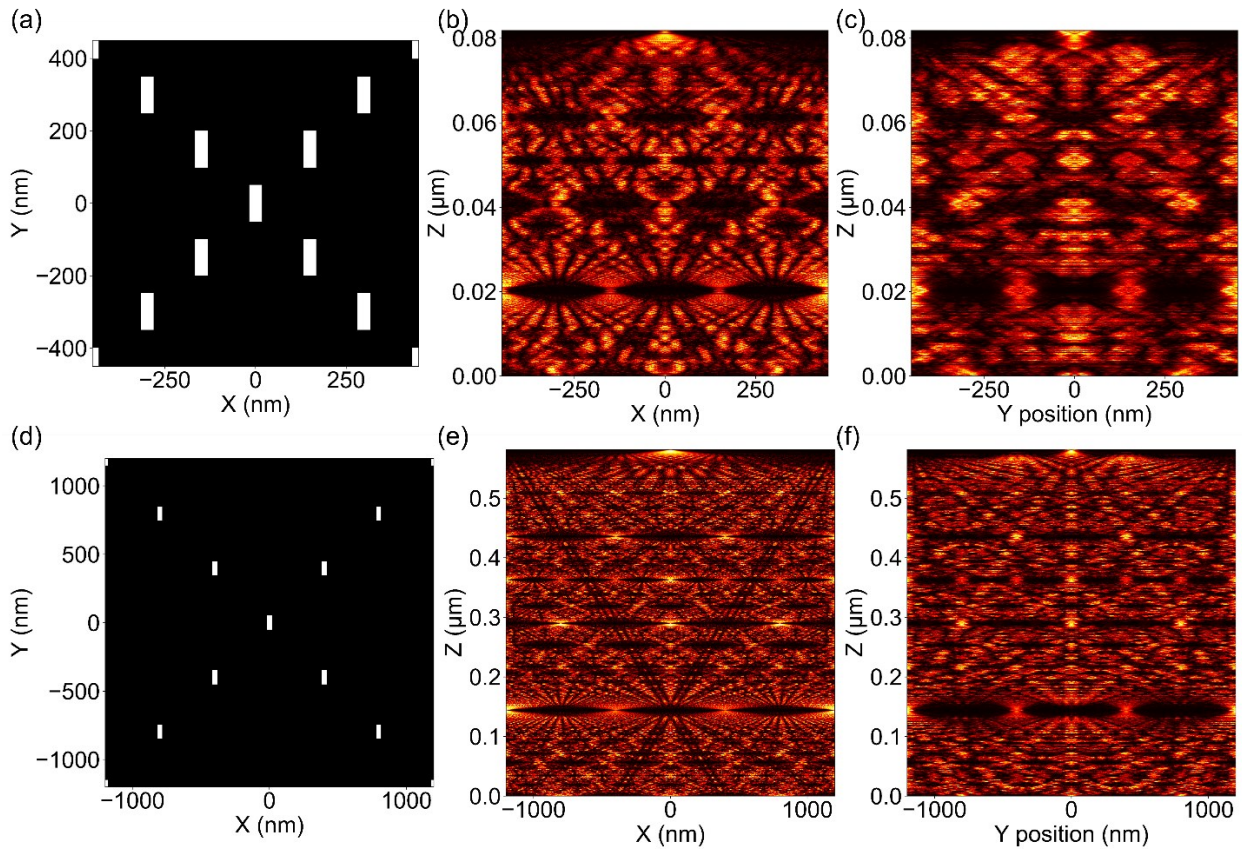

**Figure 12.** *Diffraction patterns from cross-diagonal nanorod grating configurations with varying periodicity.* (a) Schematic of cross-diagonal nanorod arrangement with 150 nm spacing. (b, c) Resulting X–Z and Y–Z probe field intensity distributions reveal dense diagonal interference fringes interlaced with irregular speckle noise, stemming from the high spatial frequency and close proximity of apertures. (d) Cross-diagonal nanorod array with increased spacing of 400 nm. (e, f) Corresponding X–Z and Y–Z slices show markedly improved pattern clarity and reduced noise, with coherent diagonal fringes and emerging Talbot-like carpets indicating partial self-imaging.

The observed diffraction behavior demonstrates strong sensitivity to aperture periodicity, fill factor, orientation, and spatial distribution, all of which can be precisely programmed using needle-beam-based structured illumination. The transition from disordered to highly regular Talbot self-imaging patterns underscores the potential of nanorods-based metastructures for reconfigurable near-field optics, including programmable diffraction gratings, beam shaping devices, and dynamically tunable photonic metasurfaces. Through structured activation alone—without physically rearranging the nanorods—it is possible to modulate complex optical functionalities, paving the way for applications in holography, light routing, and on-chip optical computing.

## 5. Conclusions

In this study, we have demonstrated a versatile, high-resolution method for voxel-scale quantum state control in dense nanorod ensembles through the use of reconfigurable needle-beam arrays. By combining spatially structured illumination with tunable optical standing waves, we achieved selective excitation of individual *InAs* nanorods embedded within a *GaAs* host, enabling site-specific manipulation of excitonic and biexcitonic transitions. This approach, validated through comprehensive multiscale simulations—including finite-element eigenstate modeling, variational Monte Carlo calculations, and Lindblad master equation analysis.

A key outcome of this work is the realization of controllable EIT in a three-level ladder-type system under variable nanorod activation conditions. By modulating the activation ratio—i.e., the fraction of nanorods illuminated by the control field—we demonstrated deterministic control over both absorption spectra and refractive index dispersion, with EIT bandwidths enhanced by a factor of six and photoluminescence peak shifts up to 80 meV. Furthermore, structured illumination enabled the creation of spatially distinct sub-populations within the ensemble, effectively acting as independent quantum nodes governed by localized optical fields.

Importantly, our platform also functions as a programmable diffractive metastructure. We demonstrated the emergence of near-field Talbot self-imaging carpets in periodically activated nanorod arrays, confirming that structured activation can serve not only to control quantum states but also to modulate the spatial phase and amplitude of transmitted probe fields. These diffraction carpets act as a direct diagnostic of the activation geometry, offering a route toward self-verifying light–matter interaction maps.

The optical crosstalk minimization, volumetric addressability, and dynamic reconfigurability inherent in the needle beam system offer new functionality for several photonics applications. In quantum information, voxel-resolved activation enables precise qubit initialization, coupling, and readout—capabilities essential for scalable architectures. In integrated nanophotonics, real-time tuning of absorption and refractive index enables on-chip optical delay lines, beam steering, and reconfigurable wavefront shaping. Additionally, this platform lays the groundwork for the creation of active metamaterials, slow-light components, and nanoscale sensors with programmable optical responses.

## ACKNOWLEDGMENTS

This work was supported by the RA MESCS Higher Education and Science Committee (Research project № 23RL-1B004).